\documentclass[conference,10pt]{IEEEtran}
\DeclareUnicodeCharacter{0301}{\'{a}}
\usepackage{amssymb,amsmath,amsfonts,bm,epsfig,graphicx,theorem,latexsym}
\usepackage{rotating,setspace,latexsym,epsf,color,subfigure}
\usepackage{cite,authblk,comment,caption}
\usepackage{hyperref}
\hypersetup{
    colorlinks=true,
   linkcolor=magenta,
    filecolor=magenta,      
    urlcolor=magenta,
    citecolor=magenta,
    pdftitle={Overleaf Example},
    pdfpagemode=FullScreen,
    }

\newtheorem{Theorem}{Theorem}

\newtheorem{Lemma}{Lemma}

\def \n2{{N_0 \over 2}}

\def \h5{\hspace{0.5in}}

\begin{document}
\IEEEoverridecommandlockouts
\pagestyle{empty}
\pagestyle{plain}

\title{How Costly Was That (In)Decision?}
\author{Anonymous author(s)}
\author{Peng Zou$^{1}$, Ali Maatouk$^{2}$, Jin Zhang$^{1}$, and Suresh Subramaniam$^{1}$ \\
	\normalsize $^{1}$ECE Department, 
	George Washington University, 
	Washington DC, 20052, USA\\
		\normalsize $^{2}$Paris Research Center, Huawei Technologies, Boulogne-Billancourt, France \\
	\normalsize $^{1}$\{{\it pzou94,zhangjin,suresh\}@gwu.edu}\ \ $^{2}$\{{\it ali.maatouk@huawei.com\}}}
\maketitle 
\begin{abstract}

In this paper, we introduce a new metric, named Penalty upon Decision (PuD), for measuring the impact of communication delays and state changes at the source on a remote decision maker. Specifically, the metric quantifies the performance degradation at the decision maker's side due to delayed, erroneous, and (possibly) missed decisions. We clarify the rationale for the
metric and derive closed-form expressions for its average in M/GI/1 and M/GI/1/1 with blocking settings. Numerical results are then presented to support our expressions and to compare the infinite and zero buffer regimes. Interestingly, comparing these two settings sheds light on a buffer length design challenge that is essential to minimize the average PuD.
\end{abstract}

\section{Introduction}
 The timeliness of information from continuous data streams is crucial for the proper functioning of a variety of Internet of Things (IoT) and edge computing applications, such as sensor networks, cognitive radio, and vehicular communication networks. For this reason, a metric named the Age of Information (AoI) has been proposed in \cite{Kaul2012} to measure the timeliness of status updates in various communication or remote-controlled systems. This metric is particularly useful in applications where timely updates are essential for the decision-making process, such as autonomous driving systems, remote surgery systems, and industrial IoT. Since its introduction in \cite{Kaul2012}, the AoI has attracted a significant amount of research attention and has been analyzed and optimized in numerous single-source and single-server settings (e.g., see \cite{costa2016age,najm2016age,Sun2016,inoue2018general,Maatouk2018,Champati2019,zou2020}).


Although the AoI metric is widely recognized as an effective means of assessing the freshness of information in time-sensitive applications, its narrow focus limits its utility. Precisely, the AoI metric only measures the timeliness of information and fails to consider other critical factors that can impact system performance, such as information mismatch between the transmitter and receiver. In contexts such as remote estimation, this limitation can result in suboptimal outcomes. 
To address this, a new metric called the Age of Incorrect Information (AoII) was proposed in \cite{Maatouk2020}. Concretely, the AoII takes into account the mismatch of information between the transmitter and the receiver in its formulation and quantifies the negative impact that this mismatch has on the system's performance over time. Numerous research problems related to the AoII have been since investigated in the literature (e.g., \cite{Kriouile2021,Chen2021,Saha2022,Maatouk2022,Liu2022}). Another limitation of the AoI metric is that it assumes the receiver is always interested in obtaining fresh information. However, in systems where a decision-maker is present, fresh information is only needed at specific decision-making points. To address this, the Age upon Decision (AuD) metric was introduced in the literature \cite{Dong2018}. Similarly, the Query Age of Information (QAoI) metric was proposed to  measure the timeliness of updates at the exact moments when the destination node will utilize the incoming information \cite{Ildiz2022}. 
Additionally, recent studies have introduced goal-oriented measures that capture the costs associated with decisions, such as the cost of actuation error \cite{9551200}. 


While the AoII metric measures the impact of information mismatch between the transmitter and receiver on the system's performance, it implicitly assumes that this mismatch is always harmful. However, this is not necessarily the case when a decision-maker is present at the receiver's side, as this harm can be mitigated due to state transitions at the source before the decision instant. Meanwhile, the AuD and QAoI metrics fail to capture the system's performance degradation when a decision-maker makes wrong decisions as a result of an information discrepancy between the transmitter and receiver. 
Furthermore, the cost of actuation error metric, while effective in capturing the effect of state mismatch at the decision instant, falls short in capturing the impact of missed decisions or correct but delayed decisions. As a result, there remains a need for further research and development of metrics that can account for a broader range of performance factors and facilitate more accurate and comprehensive evaluations of system performance at all possible decision instants. 
In this paper, we propose a new metric called Penalty upon Decision (PuD) for a multi-state information source with a decision-maker at the receiver side. The PuD captures the system's performance loss at all decision instants, considering delayed, erroneous, and (possibly) missed decisions. To that end, our contributions are summarized below:

\begin{itemize}
    \item  We begin by explaining the reasoning behind the PuD metric and outlining its general form. This form considers all the relevant parameters that can be customized to suit the specific needs of the application at hand.
    \item Next, we examine a specific case of the PuD metric and develop closed-form expressions for its average in both the M/GI/1 and M/GI/1/1 with blocking systems. Our calculations are based on stochastic analysis of the expected PuD that considers all the possible
events.
    \item Finally, we present our numerical implementations that demonstrate the accuracy of our derived closed-form expressions and provide comparisons of different queuing schemes and parameter settings in terms of average PuD. Our results reveal a performance trade-off between the M/GI/1/1 and M/GI/1 settings, highlighting the importance of buffer length design in minimizing the PuD.
\end{itemize}
  The rest of the paper is organized as follows. We present our system model in Section~\ref{sec:sys}. In Section~\ref{sec:propos}, the rationale for the metric is provided and an analysis of the measure is conducted in Sections~\ref{sec:MG1} and \ref{sec:MG11}. Finally, the numerical results and the ensuing discussions are laid out in Section~\ref{sec:Num} while Section~\ref{sec:Con} concludes the paper.

\section{System Model}\label{sec:sys}
\begin{figure}[!t] 	\centering\includegraphics[width=3.4in]{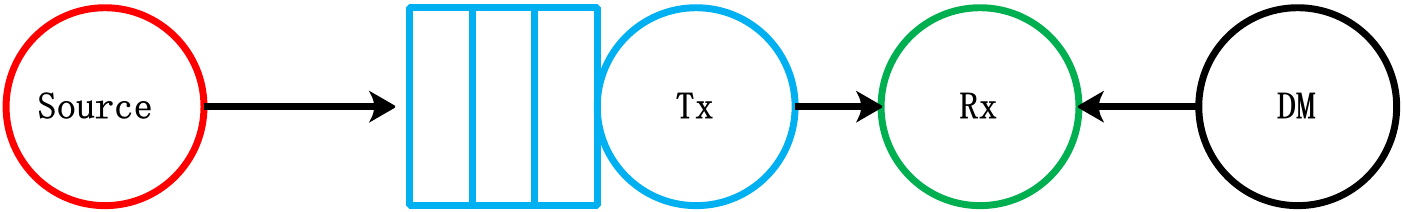}
	\caption{Illustration of our system where status updates packets arrive at a single server transmission queue with a decision maker at the receiver side.} \label{Sys}
 \vspace{-15pt}
\end{figure}
In our paper, we examine a point-to-point communication system with a single server sending status updates from a source to a receiver where a decision maker is located. An illustration of the system is shown in Fig. \ref{Sys}. We consider that the source generates packets containing information about a time-varying discrete stochastic process. We use $s_i$ to denote the state recorded by the $i$th packet, where $s_i\in \mathbb{S}$ and $\mathbb{S}$ is the state space to which $s_i$ belongs. 
%
%
We will also assume that the source only generates a new update packet when a state transition happens, and these packets arrive at the server as a Poisson process with rate $\lambda$. When an update packet is received, the decision maker makes a decision based on the state recorded by the packet after a waiting time $D$ (which we assume to be $0$ for simplicity). In other words, we consider that a decision is made immediately when the packet is received. 

In this paper, we cover the M/GI/1 and M/GI/1/1 queuing systems. In the M/GI/1 system, packets are stored in an infinite buffer and are served on a first-come-first-serve basis. In contrast, the M/GI/1/1 system has no buffer, and packets arriving while the server is busy are discarded. We assume that the service times of the packets are i.i.d. (independent and identically distributed) and follow a general distribution $f_T(t)$, $t \geq 0$. We use $\mathcal{M}_{T}(\gamma)$ to denote the moment generating function of the service time distribution evaluated at $\gamma$
\begin{align}
	\mathcal{M}_{T}(\gamma) \triangleq \mathbb{E}[e^{\gamma T}],
\end{align}
where we are interested in $\gamma \geq 0$. To use in the ensuing analysis, we also define the following quantity
\begin{align}
	\mathcal{M}_{(T,n)}(\gamma) \triangleq \mathbb{E}[T^ne^{\gamma T}],
\end{align}
where $\mathcal{M}_{(T,n)}(\gamma)$ denotes the $n$th derivative of the moment generating function of $T$  with respect to $\gamma$. Note that $\mathbb{E}[T^n]$ can be obtained from  $\mathcal{M}_{(T,n)}(\gamma)$  by evaluating $\mathcal{M}_{(T,n)}$ at $\gamma=0$.

\section{ Penalty upon Decision}\label{sec:propos}
  With the system model clarified, we delve into the details of the proposed PuD metric. We assume that for each state's transition event that takes place at the source, a corresponding decision is supposed to be made at the receiver side. With this in mind, we distinguish between three types of decisions:
  \begin{itemize}
\item \textbf{Correct decisions}:
If the state used for a decision is identical to the instantaneous state at the source, then this decision will be considered to be correct.
\item \textbf{Incorrect decisions}: A decision is said to be incorrect if the state used for the decision is different from the instantaneous state at the source.
\item \textbf{Missed decisions}: If a packet is dropped by the system, the corresponding decision is said to be a missed decision by the system.
\end{itemize}
To motivate our PuD metric, we consider an autonomous driving system in which a sensor is detecting the distance between the car and other vehicles in front. Based on the states observed by the monitor, a speed-up or slow-down decision will be made by the decision maker. In this system, it is clear that correct but delayed decisions, incorrect decisions, and missed decisions have the potential to cause an accident. Therefore, an appropriate penalty should be considered for these types of decisions to evaluate the system's performance.

With this in mind, we assume that if a decision is correct, the penalty will be related to the delay between the instant when this packet is generated and the instant when the decision is made. This delay reflects the deviation from an ideal scenario, where the decision maker would have immediately responded to the new state of the system. To that end, we define the Penalty upon Correct Decision (PuCD) for packet $i$ as
\begin{align}\label{firstsigma}
    \sigma_{i,C}=f_C(\Delta_i), 
\end{align}
where $f_C(\cdot)$ is a non-decreasing function and $\Delta_i$ represents the delay in making a decision for packet $i$. On the other hand, if an incorrect decision is made for packet $i$, we let 
\begin{align}
    \sigma_{i,I}=f_I(\Delta_i,\delta_s)
\end{align}
denote the Penalty upon Incorrect Decision (PuID), where $f_I(\Delta_i,\delta_s)$ is a non-decreasing function of the delay and the difference $\delta_s$ between the packet's state and the source's state. The intuition behind such an expression is that as the difference $\delta_s$ grows, a higher penalty should be incurred by the system. Lastly, if packet $i$ is dropped in the queuing phase due to the server being busy, we define the Penalty upon Missed Decision (PuMD) as
\begin{align}\label{lastsigma}
    \sigma_{i,M}=f_M(t,\delta_s,n_{i}),
\end{align}
where $n_i$ denotes the sequential position of the dropped packet $i$ relative to all other packets dropped during the server's current busy period. There exists a wide variety of choices for the above penalty functions depending on the application at hand. In this paper, we adopt the following expressions:
\begin{align}\label{PuD_Fun}
	\sigma_{i,C}=\Delta_i,\ 
	\sigma_{i,I}=\Delta_i+|\delta_s|\Delta_i^{|\delta_s|+1},\
	\sigma_{i,M}=n_{i}r_{n_{i}}^{(|\delta_s|+1)},
\end{align}
where $r_{n_{i}}$ represents the time interval between the arrival of the dropped packet $i$ in the system and the time the decision-maker makes a decision upon reception of the packet in service by the system. Based on these expressions, one can see that the PuCD is nothing but the delay of the packet. For the PuID, an incorrect decision incurs an additional penalty term that is added to the delay. This additional penalty is designed to reflect the greater impact of an incorrect decision compared to a correct one. Note that as $\delta_s \to 0$, PuID approaches PuCD, reflecting the desired outcome. In the case of missed decisions, we adopt a hypothetical scenario where the dropped packet is redirected to a ``virtual" server instead of being discarded due to the server being busy. The decision-making entity will then make a decision upon reception of the actual packet in service. In this context, $r_{n_{i}}$ signifies the ``imagined" time lag between the arrival of the dropped packet and the decision maker taking a decision. With this in mind, and inspired by both the PuCD and PuID, the PuMD will depend on $r_{n_{i}}$, along with the difference between the source's state at the decision-making instant and the state recorded by the dropped packet in question. Lastly, to address the issue of missed packets that arrive close to the decision-making moment having less impact, we incorporate the index $n_i$ into the penalty calculation to give these packets added importance.

\begin{figure}[!t] 	\centering\includegraphics[width=3.2in]{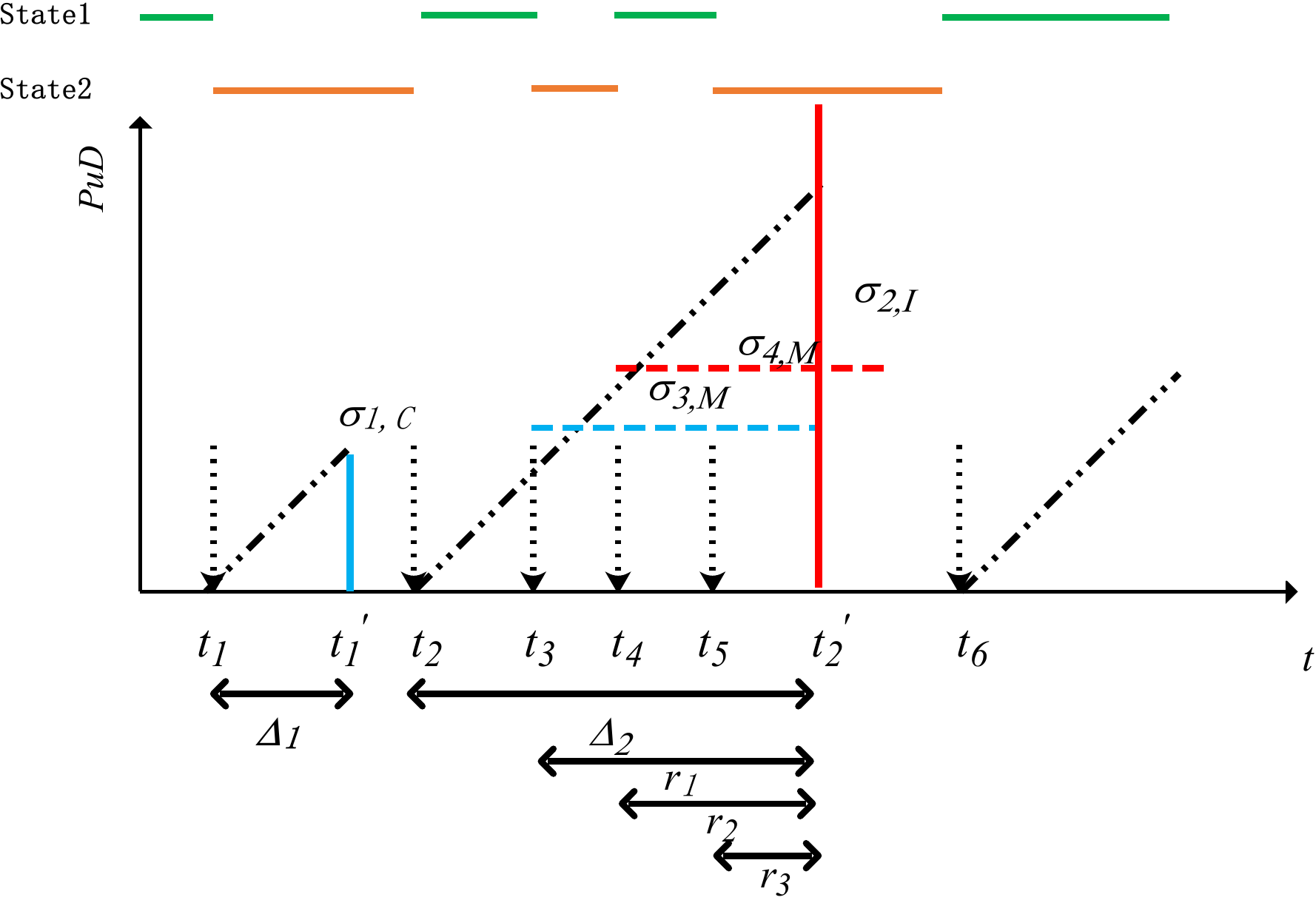}
	\caption{Evolution of the PuD in an M/GI/1/1 system.} 
 \label{Evol_AoD}
 \vspace{-20pt}
\end{figure}

To further clarify our proposed decision penalty metrics, we illustrate their evolution with an example in Fig. \ref{Evol_AoD} for an M/GI/1/1 system. 
In the remainder of the paper, we will consider that $\mathbb{S}=\{1,2\}$. The initial state of the source is state 1 and the state changes to state 2 at time $t_1$ where an update packet recording the information of state 2 is generated and served immediately by the transmitter. The transmission finishes and the decision is made at time instant $t_1'$. Therefore, it is a correct decision and the penalty for this packet equals the delay $\Delta_1$. Then, at time $t_2$, the state of the source changes to state 1, and packet 2 is served by the transmitter. At times $t_3$, $t_4$, and $t_5$, new packets are generated due to the state transition and are dropped since the transmitter is busy. Therefore, those packets are missed packets and their penalties will be related to the state of the source when the decision is made for packet 2. At time $t_2'$, the transmission finishes, and the decision is made while the source is in state 2. However, the state recorded by packet 2 is state 1. Therefore, this decision is an incorrect decision where the penalty is taken as larger than $\Delta_2$. As for packet 3, it is treated as a missed correct packet with a penalty equal to $r_1=t_2'-t_3$. On the other hand, for packet 4, it is considered a missed incorrect packet with a penalty equal to $2r_2^2$, where $r_2=t_2'-t_4$ and with the squaring accounting for $\delta_s=1$. Packet 5 is also a correct missed packet with a penalty of $3r_3$ where $r_3 = t_2' - t_5$. 

\section{M/GI/1 queue Analysis}\label{sec:MG1}
We start with an M/GI/1 system with infinite buffer size. Accordingly, no packets will be dropped, and the missed decision portion of our metric will not play a role. Since we are investigating a two-states information source, whether a decision will be correct or incorrect depends on the number of packets that have arrived during its service period. If an even number of packets arrive during a packet's service period, then the decision for the received packet will be a correct decision. Otherwise, an incorrect decision will be made for this packet. To analyze the PuD in these settings, we provide the following lemma.
  \begin{Lemma}\label{Lem1}
Let $Y=W+T$ designate the system time of the packets, where $W$ and $T$ denote the waiting time in the queue and service time, respectively. Then, the moment generating function of $Y$ is
\begin{align}
    \mathcal{M}_Y(\gamma)=\frac{-\gamma(1-\lambda\mathbb{E}[T]) \mathcal{M}_T(-\gamma)}{-\gamma-\lambda+\lambda \mathcal{M}_T(-\gamma)}.
\end{align}
	\end{Lemma}
 \begin{IEEEproof}
This Lemma can be proved by leveraging the Pollaczek-Khinchine transform
reported in \cite[Chapter~2.9]{Sztrik2012}.
\end{IEEEproof}
With the above in mind, we can derive the following theorems. 
 \begin{Theorem}\label{TMG1-P}
In an M/GI/1 system, the probabilities of correct and incorrect decisions are, respectively, equal to
\begin{align}
	p_C&=\frac{1}{2}+\frac{1}{2}M_Y(-2\lambda),\\
	p_I&=\frac{1}{2}-\frac{1}{2}M_Y(-2\lambda).
\end{align}
\end{Theorem}
	\begin{IEEEproof} The details can be found in Appendix \ref{MM1-P}.
\end{IEEEproof}
Next, using the penalties outlined in eq. (\ref{PuD_Fun}), we can obtain below the penalty upon decision, given that a correct and incorrect decision is made, respectively.
\begin{Theorem}\label{TMG1-PA}
In an M/GI/1 system, the average PuCD and PuID are, respectively, equal to
\begin{align}
	\mathbb{E}[\sigma_C]&=\frac{1}{2p_C}(\mathbb{E}[Y]+\mathcal{M}_{(Y,1)}(-2\lambda)),\\
	\mathbb{E}[\sigma_I]&=\frac{1}{2p_I}(\mathbb{E}[Y]+\mathbb{E}[Y^2]-\mathcal{M}_{(Y,1)}(-2\lambda)-\mathcal{M}_{(Y,2)}(-2\lambda)).
\end{align}
\end{Theorem}
	\begin{IEEEproof} The details can be found in Appendix \ref{MM1-PA}.
\end{IEEEproof}
In light of the above theorems, we can conclude that the total average PuD of the system, denoted by $\mathbb{E}[\sigma]$, is
\begin{equation}
    \mathbb{E}[\sigma]=\mathbb{E}[\sigma_C]p_C+\mathbb{E}[\sigma_I]p_I=\frac{1}{2}(2\mathbb{E}[Y]+\mathbb{E}[Y^2]-\mathcal{M}_{(Y,2)}(-2\lambda)).
\end{equation}

\section{M/GI/1/1 queue Analysis}\label{sec:MG11}
This section presents the theoretical analysis for an M/GI/1/1 system, where a packet will be dropped if it finds the server busy. Given this, the probability of a missed decision is equal to the probability that a packet arrives during a busy period. Note that the system's renewal structure allows us to calculate the missed decision probability as the ratio of the average busy time during one renewal cycle to the average length of a renewal cycle \cite{Gallager2011}. Therefore, the probability of missed decision can be written as
\begin{equation}
    p_M=\frac{\mathbb{E}[T]}{\frac{1}{\lambda}+\mathbb{E}[T]},
\end{equation}
where $\frac{1}{\lambda}$ is the expected idle period. Next, we investigate the probability of a correct and incorrect decision. 
 \begin{Theorem}\label{TMG11-P}
In the M/GI/1/1 system, given that a decision is made for a packet (i.e., not missed), the probabilities of correct
and incorrect decisions are given by
\begin{align}
	\Tilde{p}_C&=\frac{1}{2}+\frac{1}{2}\mathcal{M}_T(-2\lambda),\\
	\Tilde{p}_I&=\frac{1}{2}-\frac{1}{2}\mathcal{M}_T(-2\lambda).
\end{align}
\end{Theorem}
	\begin{IEEEproof} The details can be found in Appendix \ref{MG11-P}.
\end{IEEEproof}
Knowing that the above probabilities are conditioned on the event that the packet is not missed, we can conclude that the correct and incorrect decision probabilities for all the generated packets are $p_C=\Tilde{p}_C(1-p_M)$ and $p_I=\Tilde{p}_I(1-p_M)$, respectively. Next, using the penalties outlined in eq. (\ref{PuD_Fun}), we can derive the following results.
 \begin{Theorem}\label{TMG11-PA}
In the M/GI/1/1 system, the average PuCD and PuID for delivered packets are
\begin{align}
	\mathbb{E}[\sigma_C]&=\frac{1}{2\Tilde{p}_C}(\mathbb{E}[T]+\mathcal{M}_{(T,1)}(-2\lambda)),\\
	\mathbb{E}[\sigma_I]&=\frac{1}{2\Tilde{p}_I}(\mathbb{E}[T]+\mathbb{E}[T^2]-\mathcal{M}_{(T,1)}(-2\lambda)-\mathcal{M}_{(T,2)}(-2\lambda)).
\end{align}
\end{Theorem}
	\begin{IEEEproof} The details can be found in Appendix \ref{MG11-PA}.
\end{IEEEproof}
 What remains is to derive an expression of the average PuMD. To do so, we consider that a packet is being served by the system, and we condition on a given service period time $T=t$. Furthermore, we assume that $M=m$ packets arrive and are dropped during this period. Then, for the $n$th dropped packet during this service time, we define $r_n=t-X_n$, where $X_n$ is the time elapsed between the arrival of the packet in service and the arrival time of the $n$th dropped packet. Given that the PuMD depends on $r_n$, as depicted in eq. (\ref{PuD_Fun}), we characterize the distribution of $X_n$ in the following lemma. 
  \begin{Lemma}\label{Lem2}
 Given a system service period $t$ and $m$ packets arriving within this service period, the probability density function of the random variable $X_n$ is given by
 \begin{align}
 \nonumber    f_{X_n}(x_n)=&\\
     \frac{m}{t}&\binom{m-1}{n-1}\left(\frac{x_n}{t}\right)^{n-1}\left(1-\frac{x_n}{t}\right)^{m-n}, \:\: x_n\in[0,t],
 \end{align}
 for $n=1,\ldots,m$.
	\end{Lemma}
 \begin{IEEEproof} The proof's details are provided in Appendix \ref{ALem2}.
\end{IEEEproof}
Next, for every dropped packet $n$, we make the distinction between the PuMD for a missed correct decision versus the PuMD for a missed incorrect decision, denoted by $H_{\textnormal{MC}}(\cdot)$ and $H_{\textnormal{MI}}(\cdot)$ respectively. Following the definitions reported in eq. (\ref{PuD_Fun}), and considering the results of the above lemma, we have
 \begin{align}
     H_{\textnormal{MC}}(n|T=t,M=m)&=\int_{0}^{t}n(t-x_n)f(x_n)dx_n\nonumber\\ &=\left(n-\frac{n^2}{m+1}\right)t,
 \end{align}
  \begin{align}
 \nonumber    H_{\textnormal{MI}}(n|T=t,&M=m)=\int_{0}^{t}n(t-x_n)^2f(x_n)dx_n\\&=\left(n-\frac{2n^2}{m+1}+\frac{n^2(n+1)}{(m+2)(m+1)}\right)t^2.
 \end{align}
 Note that for a two-states system, whether a dropped packet will lead to a missed incorrect decision or a missed correct decision depends on the number of packets that arrive between the packet in question and the end of the service period $t$. With this in mind, and based on the indices $m$ and $n$, we use $I\in\{I_1,I_2,I_3,I_4\}$ to denote the events of whether a dropped packet leads to a missed correct or incorrect decision. The details of these events are summarized below.
 \begin{itemize}
     \item Event $I_1$: If $m$ and $n$ are even numbers, then the packet with index $n$ will be a correct missed packet.
     \item Event $I_2$: If $m$ is an even number and $n$ is an odd number, the packet with index $n$ will be an incorrect missed packet.
     \item Event $I_3$: If $m$ is an odd number and $n$ is an even number, the packet with index $n$ will be an incorrect missed packet.
     \item Event $I_4$: If $m$ and $n$ are odd numbers, the packet with index $n$ will be a correct missed packet.
 \end{itemize}
 By leveraging the above events, a step forward towards calculating the average PuMD consists of characterizing the PuMD jointly with any of the aforementioned events. Then, the law of total expectation can be applied by integrating over the whole space of $T$ and $M$ to obtain $\mathbb{E}[\sigma_{M, I=I_j}]$ for $j=1,\ldots,4$. With this in mind, we report below our results.
 \begin{figure*}[t]
\setlength{\abovecaptionskip}{-5pt}
\setlength{\belowcaptionskip}{-10pt}
\centering
\captionsetup{width=0.3\linewidth}
\begin{minipage}[t]{0.32\linewidth}
\centering\includegraphics[width=2.35in]{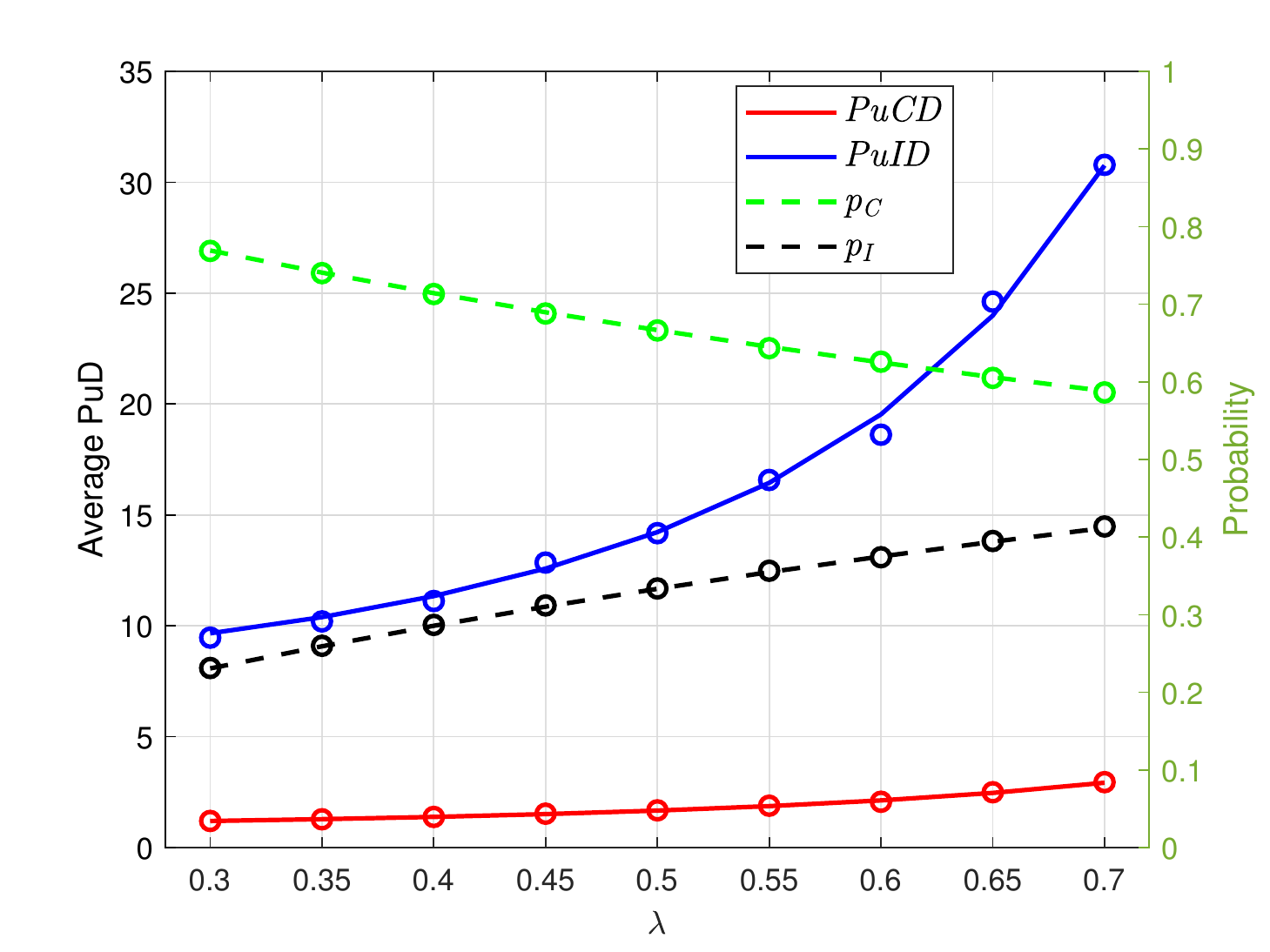}
	\caption{PuD and decision probability versus $\lambda$ in M/M/1 system with $\mu=1$.} \label{Num1}
\end{minipage}
\begin{minipage}[t]{0.3\linewidth}
\centering\includegraphics[width=2.35in]{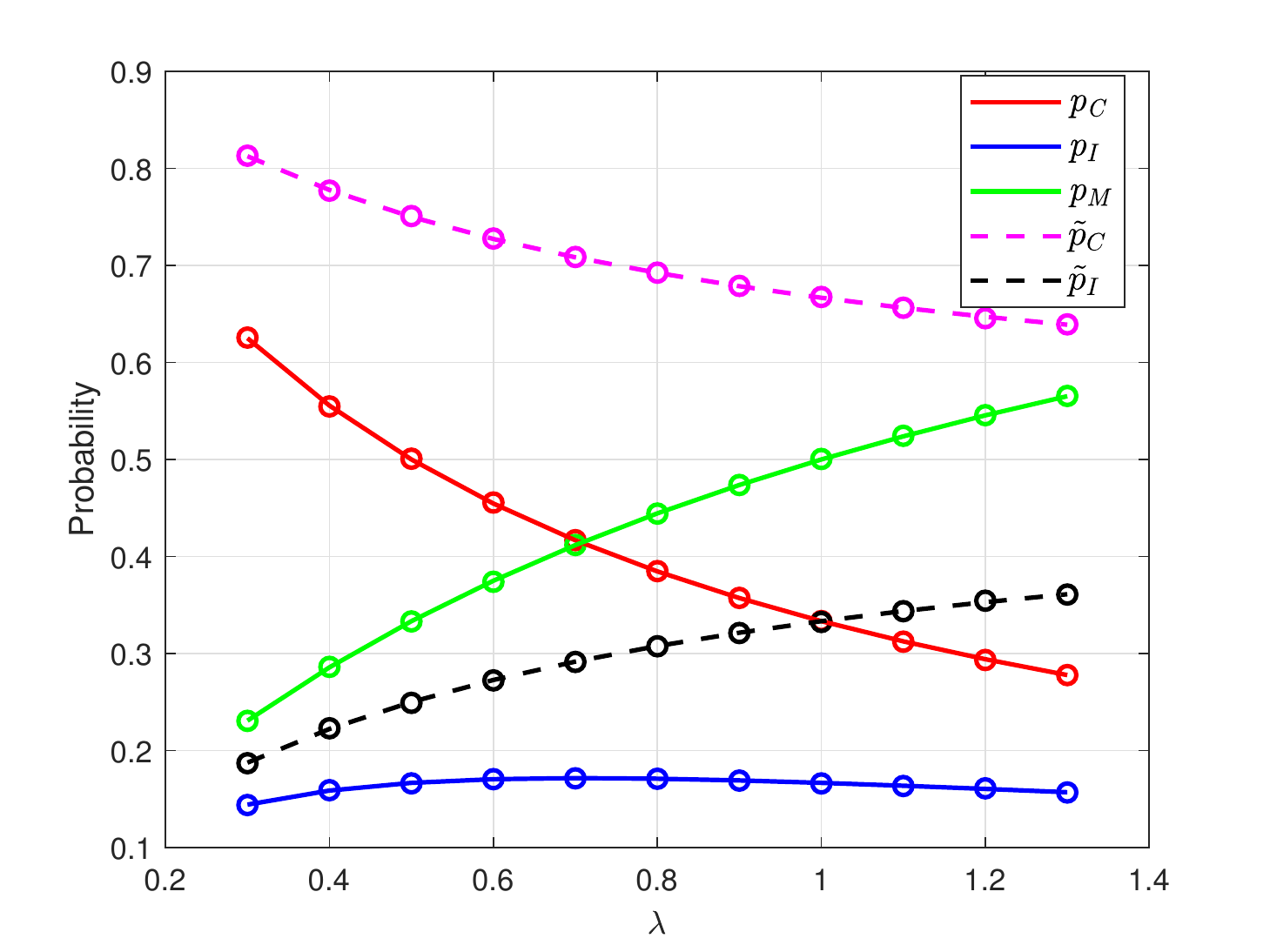}
	\caption{Decision probability versus $\lambda$ in M/M/1/1 system where $\mu=1$.} \label{Num2}
\end{minipage}
\begin{minipage}[t]{0.32\linewidth}
\centering\includegraphics[width=2.35in]{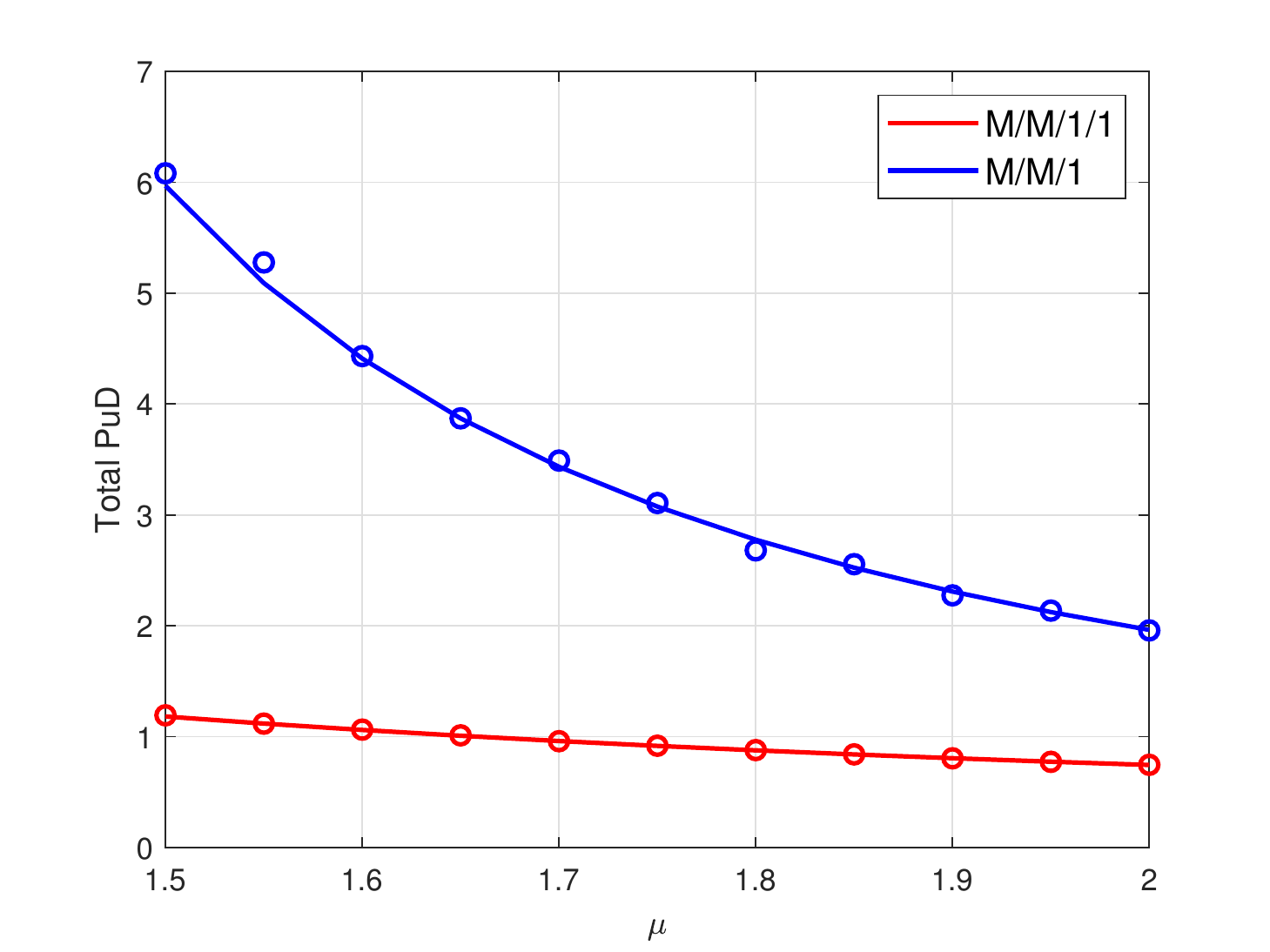}
	\caption{Total PuD versus $\mu$ in M/M/1 and M/M/1/1 systems where $\lambda=1$.} \label{Num3}
\end{minipage}
\vspace{-9pt}
\end{figure*}
  \begin{Theorem}\label{TMG11-M}
The average PuMD, when considered jointly with the event $I_1$ and given that a packet is being served by the system, has the following expression
\begin{align}
\nonumber	\mathbb{E}[\sigma_{M, I=I_1}]=&
	\frac{\lambda \mathbb{E}[T^2]}{8} 
	+ \frac{\lambda^2\mathbb{E}[T^3]}{24}
	-\frac{\lambda \mathcal{M}_{(T,2)}(-2\lambda)}{8}\\& 
	+\frac{\lambda^2\mathcal{M}_{(T,3)}(-2\lambda)}{24}.
\end{align}
\end{Theorem}
	\begin{IEEEproof} The details are provided in Appendix \ref{MG11-M}. 
\end{IEEEproof}
Due to space constraints, and given that the remaining results can be obtained by following a similar method, we report the final equations of $\mathbb{E}[\sigma_{M, I=I_j}]$ for $j=2,\ldots,4$ in Appendix \ref{MG11_MR}. Now, given that the set $\{I_1,\ldots,I_4\}$ forms a partition of the probability space, we first sum together $\mathbb{E}[\sigma_{M, I=I_j}]$ for $j=1,\ldots,4$. We then note that the results from the above theorem are based on the assumption that a packet is served by the system. Thus, we multiply the aforementioned sum by $1-p_M$, which represents the probability of a packet being served. Lastly, since the average PuMD is defined as being conditioned on a packet being missed, we divide the resulting quantity by $p_M$. In conclusion, the PuMD can be expressed as follows:
\begin{equation}
    \mathbb{E}[\sigma_M]=\left(\sum_{I_j\in\{I_1,I_2,I_3,I_4\}}\mathbb{E}[\sigma_{M,I=I_j}]\right)\frac{1-p_M}{p_M}.
\end{equation}


\section{Numerical Results}\label{sec:Num}

In this section, we provide numerical implementations to characterize the probability of the possible types of decisions along with their associated penalties. Our simulations are packet-based and were run with a total of one million packets to guarantee convergence.
In all figures, the simulation results are depicted as circles, while the theoretical expressions from our analysis are displayed as solid lines. The results show a strong match between the simulation outcomes and the theoretical expressions, confirming the validity of our analysis. With this in mind, we now delve deeper into the dynamics of the quantities involved and gain further insights into the systems under consideration.


We start with Fig. \ref{Num1}, which displays both the probabilities of the possible types of decisions and their corresponding penalties in an M/M/1 system. With a server service rate $\mu$ fixed to $1$, we vary $\lambda$ to observe how that would affect the quantities involved. As $\lambda$ increases, we can see that $p_C$ decreases while $p_I$ increases. This indicates that as $\lambda$ becomes large, the waiting time for packets in the queue becomes longer, increasing the likelihood of an incorrect decision being made. As $\lambda$ approaches $1$, both the correct and incorrect decision penalties approach $0.5$. To understand this trend, we recall that as $\lambda$ becomes close to its maximum value $1$ (beyond which the queue becomes unstable), the queuing time becomes extremely large. Therefore, by the time a packet is delivered, many others would have already arrived in the queue, indicating changes in the state of the source. At this point, the system's estimation of the source state will become equivalent to a coin toss. It's worth noting that as $\lambda$ increases, both the PuCD and PuID also rise due to the prolonged delay for each packet in the system. 


Next, in Fig. \ref{Num2}, we show the probabilities of the possible types of decisions in an M/M/1/1 system. The service rate of the server is set to $\mu=1$, and we vary $\lambda$ to observe its effect. Unlike the M/M/1 system, the M/M/1/1 queue has no buffer, resulting in dropped packets when the server is busy. As $\lambda$ increases, we can see that $p_M$ also increases since more packets are dropped in the queue. Additionally, there is a peak value for $p_I$, which is due to the fact that as $\lambda$ increases, $\Tilde{p}_I$ also increases. However, the total number of received packets decreases since a higher number of them are dropped, resulting in the showcased trend of $p_I$.

With the performance of the M/M/1 and M/M/1/1 queues depicted, our attention shifts to comparing both systems in terms of total PuD. 
We fix the arrival rate $\lambda$ to $1$, and we vary the service rate $\mu$ to observe its effect on the total PuD. As seen in Fig. \ref{Num3}, the total PuD of the M/M/1 system is consistently higher than that of the M/M/1/1 system. To understand this trend, we recall that the M/M/1 system does not exhibit a PuMD when compared to the M/M/1/1 system. Therefore, although the availability of an infinite buffer may lead to queuing delays, it also eliminates the possibility of missed decisions. The 
 results in Fig. \ref{Num3} suggest that for the PuD functions adopted in this paper and detailed in eq. (\ref{PuD_Fun}), dropping packets from the queue rather than storing them in the buffer comes at a smaller cost for the total PuD of the system. However, this is not necessarily true for any penalty function that follows the general form reported in eq. (\ref{firstsigma})-(\ref{lastsigma}). For this reason, one can conclude that a trade-off exists between the reduction of the PuMD by increasing the buffer length and the reduction of the total PuD of the system. This trade-off is closely related to the application at hand, and thus finding the optimal buffer length is an important design challenge for the minimization of the PuD. For this reason, studying this trade-off is an essential research direction for our future work. 
 Moreover, this trade-off also highlights the difference between the PuD and traditional AoI metric. While removing the buffer is always considered optimal for the AoI metric \cite{8006593}, this is not necessarily the case for the PuD metric, as a larger buffer may result in a lower PuD. This underscores the significance of introducing new performance measures, such as the PuD metric, to the literature as such introductions present new design challenges that can further improve the system's overall performance. 
 Note that as we increase the service rate $\mu$, the performance gap between the two systems decreases. This is because as the service rate approaches infinity, all packets are served immediately, making the buffer length irrelevant.

\vspace{-1pt}

\section{Conclusion}\label{sec:Con}
In this paper, we proposed the penalty upon decision metric to measure the impact of correct but delayed, incorrect, and missed decisions on a multi-state source system with a decision maker at the receiver. We provided a general form of the metric and discussed the various aspects it captures. For a specific case of the metric, we derived closed-form expressions of its average in M/GI/1 and M/GI/1/1 queueing systems with a two-state source. Our numerical implementations have shown that our closed-form expressions yield accurate results and highlighted a buffer length design challenge necessary to minimize the average penalty. 
Our future research will focus on expanding our analysis to cover multi-state sources and exploring more generalized forms of the PuD. Such an extension approach will involve utilizing the dynamics of information source transitions and examining their impact on the various penalty functions. 

\section*{Acknowledgment}
This work was supported in part by NSF grant CNS-2219180.

\bibliographystyle{ieeetr}  

\begin{thebibliography}{10}

\bibitem{Kaul2012}
S.~Kaul, R.~Yates, and M.~Gruteser, ``Real-time status: How often should one
  update?,'' in {\em 2012 Proceedings IEEE INFOCOM}, pp.~2731--2735, 2012.

\bibitem{costa2016age}
M.~Costa, M.~Codreanu, and A.~Ephremides, ``On the age of information in status
  update systems with packet management,'' {\em IEEE Transactions on
  Information Theory}, vol.~62, no.~4, pp.~1897--1910, 2016.

\bibitem{najm2016age}
E.~Najm and R.~Nasser, ``Age of information: The gamma awakening,'' in {\em
  2016 IEEE International Symposium on Information Theory (ISIT)},
  pp.~2574--2578, 2016.

\bibitem{Sun2016}
Y.~Sun, E.~Uysal-Biyikoglu, R.~Yates, C.~E. Koksal, and N.~B. Shroff, ``Update
  or wait: How to keep your data fresh,'' in {\em IEEE INFOCOM 2016 - The 35th
  Annual IEEE International Conference on Computer Communications}, pp.~1--9,
  2016.

\bibitem{inoue2018general}
Y.~Inoue, H.~Masuyama, T.~Takine, and T.~Tanaka, ``A general formula for the
  stationary distribution of the age of information and its application to
  single-server queues,'' {\em IEEE Transactions on Information Theory},
  vol.~65, no.~12, pp.~8305--8324, 2019.

\bibitem{Maatouk2018}
A.~Maatouk, M.~Assaad, and A.~Ephremides, ``The age of updates in a simple
  relay network,'' in {\em 2018 IEEE Information Theory Workshop (ITW)},
  pp.~1--5, 2018.

\bibitem{Champati2019}
J.~P. Champati, H.~Al-Zubaidy, and J.~Gross, ``On the distribution of aoi for
  the {GI/GI/1/1} and {GI/GI/1/2*} systems: Exact expressions and bounds,'' in
  {\em IEEE INFOCOM 2019 - IEEE Conference on Computer Communications},
  pp.~37--45, 2019.

\bibitem{zou2020}
P.~Zou, O.~Ozel, and S.~Subramaniam, ``Waiting before serving: A companion to
  packet management in status update systems,'' {\em IEEE Transactions on
  Information Theory}, vol.~66, no.~6, pp.~3864--3877, 2020.

\bibitem{Maatouk2020}
A.~Maatouk, S.~Kriouile, M.~Assaad, and A.~Ephremides, ``The age of incorrect
  information: A new performance metric for status updates,'' {\em IEEE/ACM
  Transactions on Networking}, vol.~28, no.~5, pp.~2215--2228, 2020.

\bibitem{Kriouile2021}
S.~Kriouile and M.~Assaad, ``Minimizing the age of incorrect information for
  real-time tracking of markov remote sources,'' in {\em 2021 IEEE
  International Symposium on Information Theory (ISIT)}, pp.~2978--2983, 2021.

\bibitem{Chen2021}
Y.~Chen and A.~Ephremides, ``Minimizing age of incorrect information for
  unreliable channel with power constraint,'' in {\em 2021 IEEE Global
  Communications Conference (GLOBECOM)}, pp.~1--6, 2021.

\bibitem{Saha2022}
S.~Saha, H.~S. Makkar, V.~B. Sukumaran, and C.~R. Murthy, ``On the relationship
  between mean absolute error and age of incorrect information in the
  estimation of a piecewise linear signal over noisy channels,'' {\em IEEE
  Communications Letters}, pp.~1--1, 2022.

\bibitem{Maatouk2022}
A.~Maatouk, M.~Assaad, and A.~Ephremides, ``The age of incorrect information:
  an enabler of semantics-empowered communication,'' {\em IEEE Transactions on
  Wireless Communications}, pp.~1--1, 2022.

\bibitem{Liu2022}
Q.~Liu, C.~Li, Y.~T. Hou, W.~Lou, J.~H. Reed, and S.~Kompella, ``Ao2i:
  Minimizing age of outdated information to improve freshness in data
  collection,'' in {\em IEEE INFOCOM 2022 - IEEE Conference on Computer
  Communications}, pp.~1359--1368, 2022.

\bibitem{Dong2018}
Y.~Dong, Z.~Chen, S.~Liu, and P.~Fan, ``Age of information upon decisions,'' in
  {\em 2018 IEEE 39th Sarnoff Symposium}, pp.~1--5, 2018.

\bibitem{Ildiz2022}
M.~E. Ildiz, O.~T. Yavascan, E.~Uysal, and O.~T. Kartal, ``Query age of
  information: Optimizing aoi at the right time,'' in {\em 2022 IEEE
  International Symposium on Information Theory (ISIT)}, pp.~144--149, 2022.

\bibitem{9551200}
N.~Pappas and M.~Kountouris, ``Goal-oriented communication for real-time
  tracking in autonomous systems,'' in {\em 2021 IEEE International Conference
  on Autonomous Systems (ICAS)}, pp.~1--5, 2021.

\bibitem{Sztrik2012}
S.~János, ``Basic queueing theory: Foundations of system performance
  modeling,'' 2016.

\bibitem{Gallager2011}
R.~G. Gallager, ``Discrete stochastic processes,'' {\em OpenCourseWare:
  Massachusetts Institute of Technology}, 2011.

\bibitem{8006593}
A.~M. Bedewy, Y.~Sun, and N.~B. Shroff, ``Age-optimal information updates in
  multihop networks,'' in {\em 2017 IEEE International Symposium on Information
  Theory (ISIT)}, pp.~576--580, 2017.

\bibitem{larson_odoni_1981}
R.~C. Larson and A.~R. Odoni, {\em Urban Operations Research}.
\newblock Prentice-Hall, 1981.

\bibitem{Gentle2010}
J.~E. Gentle, {\em Computational statistics}.
\newblock Springer, 2010.

\end{thebibliography}
\begin{appendices}
\section{Proof of Theorem \ref{TMG1-P}} \label{MM1-P} 
To prove our theorem, let us consider that a random packet's system time $Y$ is equal to $y$. Additionally, let $M$ denote the number of packets arriving during this packet's system time. Since the packets' arrivals follow a Poisson distribution, we can conclude that the probability of $m$ packets arriving during the system time period $y$ is
\begin{align}
	P(M=m|Y=y)=\frac{(\lambda y)^m e^{-\lambda y}}{m!}.
\end{align}
Given the results of Lemma \ref{Lem1}, we have that the probability of a correct decision for packets in an M/GI/1 system is:
\begin{align}
  p_C&=\mathbb{E}_{Y}[\sum_{m\ \textnormal{is even}} P(M=m|Y)] \nonumber\\
	&=\int_{0}^{\infty}\sum_{m\ \textnormal{is even}}\frac{(\lambda t)^m e^{-\lambda y}}{m!}f_Y(y)dy \nonumber\\
	&=\int_{0}^{\infty}f_Y(y)e^{-\lambda y}\sum_{m\ \textnormal{is even}}\frac{(\lambda y)^m }{m!}dy \nonumber\\
	 &=\int_{0}^{\infty}f_Y(y)e^{-\lambda y}\cosh(\lambda y)dy \nonumber\\
	&=\frac{1}{2}+\frac{1}{2}\mathcal{M}_Y(-2\lambda).
\end{align}
Now, given that an incorrect decision occurs when an odd number of packets arrive during the packet's service time, we can conclude that the probability of an incorrect decision is:
\begin{align}
p_I&=\mathbb{E}_{Y}[\sum_{m\ \textnormal{is odd}} P(M=m|Y)] \nonumber\\
&=\int_{0}^{\infty}f_Y(y)e^{-\lambda y}\sum_{m\ \textnormal{is odd}}\frac{(\lambda y)^m }{m!}dy \nonumber\\
&=\int_{0}^{\infty}f_Y(y)e^{-\lambda y}\sinh(\lambda y)dy \nonumber\\
  &=\frac{1}{2}-\frac{1}{2}M_Y(-2\lambda).
\end{align}
\section {Proof of Theorem \ref{TMG1-PA}} \label{MM1-PA} 
To derive the average PuCD, we recall that the PuCD is defined as the penalty upon decision, given that a packet delivery leads to a correct decision. Therefore, knowing that the packet delay is nothing but its system time $Y$, we can follow similar arguments to those in Appendix \ref{MM1-P} to conclude 
\begin{align}
   \mathbb{E}[\sigma_C]&=\mathbb{E}_{Y}[\sum_{m\ \textnormal{is even}}YP(M=m|Y)]/p_C \nonumber\\
 &=\frac{1}{p_C}\int_{0}^{\infty}yf_Y(y)e^{-\lambda y}\cosh(\lambda y)dy \nonumber\\
 &=\frac{1}{2p_C}(\mathbb{E}[Y]+\mathcal{M}_{(Y,1)}(-2\lambda)).
 \end{align}
Likewise, we can deduce that the average PuID is
  \begin{align}
  \mathbb{E}[&\sigma_I]=\mathbb{E}_Y[\sum_{m\ \textnormal{is odd}}(Y+Y^2)P(M=m|Y)]/p_I \nonumber\\
 &=\frac{1}{p_I}\int_{0}^{\infty}(y+y^2)f_Y(y)e^{-\lambda y}\sinh(\lambda y)dy \nonumber\\
 &=\frac{1}{2p_I}(\mathbb{E}[Y]+\mathbb{E}[Y^2]-\mathcal{M}_{(Y,1)}(-2\lambda)-\mathcal{M}_{(Y,2)}(-2\lambda)).
 \end{align}
 
 \section {Proof of Theorem \ref{TMG11-P}} \label{MG11-P} 
 Given that the probability density function of $T$ is $f_T(t)$, we can obtain the following expression for the probability of correct decision for received packets (i.e., not missed).
\begin{align}
  \Tilde{p}_C&=\mathbb{E}_T[\sum_{m\ \textnormal{is even}}P(M=m|T)] \nonumber\\
&=\int_{0}^{\infty}\sum_{m\ \textnormal{is even}}\frac{(\lambda t)^m e^{-\lambda t}}{m!}f_T(t)dt \nonumber\\
 &=\int_{0}^{\infty}f_T(t)e^{-\lambda t}\cosh(\lambda t)dt \nonumber\\
	&=\frac{1}{2}+\frac{1}{2}\mathcal{M}_T(-2\lambda).
\end{align}
 Similarly, we can obtain the following expression for the probability of incorrect decision for the received packets.
 \begin{align}
  \Tilde{p}_I&=\mathbb{E}_T[\sum_{m\ \textnormal{is odd}}P(M=m|T)] \nonumber\\
   &=\int_{0}^{\infty}f_T(t)e^{-\lambda t}\sinh(\lambda t)dt \nonumber\\
     &=\frac{1}{2}-\frac{1}{2}\mathcal{M}_T(-2\lambda).
 \end{align}
 
 \section {Proof of Theorem \ref{TMG11-PA}} \label{MG11-PA} 
 In an M/GI/1/1 queue, the delay for a packet's decision to be taken will be its service time. Therefore, the expected PuCD, given that the packet is not missed, can be derived as follows.
 \begin{align}
    \mathbb{E}[\sigma_C]&=\mathbb{E}_T[\sum_{m\ \textnormal{is even}}TP(M=m|T)]/\Tilde{p}_C \nonumber\\
 &=\frac{1}{\Tilde{p}_C}\int_{0}^{\infty}tf_T(t)e^{-\lambda t}\cosh(\lambda t)dt \nonumber\\
 &=\frac{1}{2\Tilde{p}_C}(\mathbb{E}[T]+\mathcal{M}_{(T,1)}(-2\lambda)).
 \end{align}
Likewise, we can derive the expected PuID hereby.
  \begin{align}
  \mathbb{E}[\sigma_I]&=\mathbb{E}_{T}[\sum_{m\ \textnormal{is odd}}(T+T^2)P(M=m|T)]/\Tilde{p}_I \nonumber\\
 &=\frac{1}{\Tilde{p}_I}\int_{0}^{\infty}(t+t^2)f_T(t)e^{-\lambda t}\sinh(\lambda t)dt \nonumber\\
 &=\frac{1}{2\Tilde{p}_I}(\mathbb{E}[T]+\mathbb{E}[T^2]-\mathcal{M}_{(T,1)}(-2\lambda)-\mathcal{M}_{(T,2)}(-2\lambda)).
 \end{align}

 \section{Proof of Theorem \ref{Lem2}}\label{ALem2}

\begin{figure}[!ht]
\centering
\includegraphics[width=.64\linewidth]{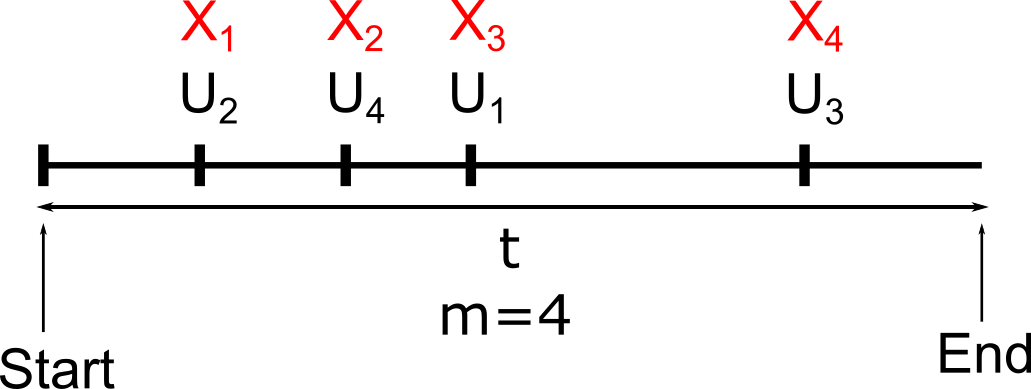}
\caption{Illustration of the packets' arrival.}
\label{proofillustration}
\end{figure}

  To prove our theorem, let us first recall that we assumed that the server was busy, and we conditioned on a given service period time $T=t$. Also, we considered that $m$ packets arrived during this service period. Given that the packets' arrival follows a Poisson distribution, we can conclude that the arrival times of these packets within the service period $T$ are independent and uniformly distributed on $[0,t]$ \cite[Chapter~2.12]{larson_odoni_1981}. Hence, let $U_1,\ldots,U_m$ denote the unordered set of these arrival times. Next, we recall that our goal is to characterize the distribution of $X_n$ for $n=1,\ldots,m$. By definition, the set $X_1,\ldots,X_m$ denotes an ordered set of the random variables $U_1,\ldots,U_m$. Particularly, 
  \begin{equation}
      X_1=\min\{U_1,\ldots,U_m\},
  \end{equation}
    \begin{equation}
      X_m=\max\{U_1,\ldots,U_m\}.
  \end{equation}
  An illustration is provided in Fig. \ref{proofillustration}. This brings us to the notion of order statistics. In essence, the $n$th order statistic of a statistical sample is equal to the $n$th smallest value. Therefore, the problem boils down to characterizing the order statistics of a set of independent uniform random variables. To do so, let us define the normalized uniform random variables $\hat{U}_n=U_n/t$ and the normalized random variables $\hat{X}_n=X_n/t$. By leveraging the results in \cite{Gentle2010}, we can conclude that the probability distribution function of $\hat{X}_n$ is
  \begin{align}\label{standardbeta}
      f_{\hat{X}_n}(\hat{x})=m\binom{m-1}{n-1}\hat{x}^{n-1}(1-\hat{x})^{m-n}, \quad \hat{x}\in[0,1].
  \end{align}
  In other words, the random variable $\hat{X}_n$ follows a beta distribution $\beta(n,m-n+1)$ for $n=1,\ldots,m$. Now, given that $X_n=t\hat{X}_n$, we can conclude that
  \begin{equation}
      f_{X_n}(x_n)=\frac{1}{t} f_{\hat{X}_n}(\frac{x_n}{t}), \quad x_n\in[0,t].
  \end{equation}
  Then, using eq. (\ref{standardbeta}), we can
conclude our proof.
\vspace{-5pt}
\vspace{60pt}
 \section {Proof of Theorem \ref{TMG11-M}} \label{MG11-M} 
To obtain the PuMD when considered jointly with event $I_1$, we first note that 
 \begin{align}
\nonumber\mathbb{E}[&\sigma_{M,I=I_1}]=\\ &\mathbb{E}_T[\sum_{m\ \textnormal{even}}\sum_{\substack{n\ \textnormal{even} \\ n\leq m}}H_{\textnormal{MC}}(n|T,M=m)P(M=m|T)].
 \end{align}
 By letting $m=2x$, $n=2y$, we obtain
 \begin{align}\label{EI1}
 \nonumber    \mathbb{E}&[\sigma_{M,I=I_1}]=\\
\nonumber     &\int_{0}^{\infty}\sum_{x=0}^{\infty}\sum_{y=0}^{x}\left(2y-\frac{4y^2}{2x+1}\right)tP(M=2x|T=t)f_T(t)dt\\
\nonumber     =&\int_{0}^{\infty}\sum_{x=0}^{\infty}(x(x+1)-\frac{2}{3}x(x+1))P(M=2x|T=t)tf_T(t)dt\\
=&\int_{0}^{\infty}\sum_{x=0}^{\infty}(x^2+x)\frac{(\lambda t)^{2x} e^{-\lambda t}}{(2x)!}\frac{t}{3}f_T(t)dt.
 \end{align}
 To simplify eq. (\ref{EI1}), we present below an analysis of the terms included in the equation:
 \begin{align}
     \sum_{x=0}^{\infty}x\frac{(\lambda t)^{2x} }{(2x)!}&=\sum_{x=0}^{\infty}\frac{\lambda t}{2}\frac{(\lambda t)^{2x-1} }{(2x-1)!}=\frac{\lambda t}{2}\sinh(\lambda t) \nonumber\\
\nonumber     \sum_{x=0}^{\infty}x^2\frac{(\lambda t)^{2x}}{(2x)!}&=\sum_{x=0}^{\infty}\frac{2x(2x-1)+2x}{4}\frac{(\lambda t)^{2x} }{(2x)!}\\
\nonumber&=\sum_{x=0}^{\infty}\left[\frac{(\lambda t)^{2}}{4}\frac{(\lambda t)^{2x-2}}{(2x-2)!}+\frac{x}{2}\frac{(\lambda t)^{2x} }{(2x)!}\right]\\
&=\frac{(\lambda t)^{2}}{4}\cosh(\lambda t)+\frac{\lambda t}{4}\sinh(\lambda t).
 \end{align}
By leveraging the above results, we can obtain below the final expression for the PuMD jointly with event $I_1$
\begin{align}
	\mathbb{E}&[\sigma_{M,I=I_1}]= \nonumber\\
	&\int_{0}^{\infty}\left[\frac{\lambda t^2}{8} + \frac{\lambda^2 t^3}{24}-\frac{\lambda t^2e^{-2\lambda t}}{8}+\frac{\lambda^2t^3e^{-2\lambda t}}{24}\right]f_T(t)dt \nonumber\\
	=&\frac{\lambda \mathbb{E}[T^2]}{8} + \frac{\lambda^2 \mathbb{E}[T^3]}{24}-\frac{\lambda \mathcal{M}_{(T,2)}(-2\lambda)}{8}+\frac{\lambda^2\mathcal{M}_{(T,3)}(-2\lambda)}{24}.
\end{align}

\section{Average PuMD Jointly with Events $I_2$, $I_3$ and $I_4$.}
\label{MG11_MR}
By following similar methods to those reported in Appendix \ref{MG11-M}, we can obtain the ensuing expressions of $\mathbb{E}[\sigma_{M, I=I_j}]$ for $j=2,\ldots,4$:
\begin{align}
    \nonumber	\mathbb{E}[\sigma_{M, I=I_2}]=&
	-\frac{\mathbb{E}[T]}{16\lambda}
	+ \frac{\mathbb{E}[T^2]}{16}
	 + \frac{\lambda\mathbb{E}[T^3] }{12}
	 + \frac{\lambda^2\mathbb{E}[T^4]}{48}\\&\nonumber
	 +\frac{\mathcal{M}_{(T,1)}(-2\lambda) }{16\lambda}
	 + \frac{\mathcal{M}_{(T,2)}(-2\lambda)}{16} \\&
	 - \frac{\lambda \mathcal{M}_{(T,3)}(-2\lambda) }{12}
 + \frac{\lambda^2\mathcal{M}_{(T,4)}(-2\lambda)}{48},
 \end{align}

\begin{align}
\mathbb{E}[\sigma_{M, I=I_3}]=
	&\frac{1}{16\lambda^2}
	-\frac{\mathbb{E}[T] }{16\lambda } 
	-\frac{\mathbb{E}[T^2]}{16} 
	+ \frac{\lambda\mathbb{E}[T^3]}{12}
	+\frac{\lambda^2\mathbb{E}[T^4] }{48}\nonumber\\&
	-\frac{\mathcal{M}_{T}(-2\lambda)}{16\lambda^2} 
- \frac{\mathcal{M}_{(T,1)}(-2\lambda) }{16\lambda}
	+\frac{\mathcal{M}_{(T,2)}(-2\lambda) }{16}\nonumber\\& 
	+ \frac{\lambda \mathcal{M}_{(T,3)}(-2\lambda)}{12}
	 - \frac{\lambda^2\mathcal{M}_{(T,4)}(-2\lambda)}{48},
  \end{align}
\begin{align}	\mathbb{E}[\sigma_{M, I=I_4}]=&
	\frac{\mathbb{E}[T]}{8}
	+ \frac{\lambda \mathbb{E}[T^2]}{8}
	+ \frac{\lambda^2\mathbb{E}[T^3]}{24} 
	- \frac{\mathcal{M}_{(T,1)}(-2\lambda)}{8}  \nonumber\\&
	+ \frac{\lambda \mathcal{M}_{(T,2)}(-2\lambda)}{8} 
	- \frac{\lambda^2\mathcal{M}_{(T,3)}(-2\lambda)}{24}.
\end{align}
\end{appendices}

\end{document}